\documentstyle[aps,epsf,times,float,graphicx]{revtex}
\DeclareGraphicsExtensions{.eps,.ps} 
 
\begin{document}
\draft
\twocolumn[\hsize\textwidth\columnwidth\hsize\csname@twocolumnfalse\endcsname 
\title{Transport Properties Calculation in the Superconducting State for a 
Quasi-Twodimensional System } 
\author{C. P. Moca$^{1,2}$ and E. Macocian$^{2}$} 
\address{
$^1$ Department of Physics, Notre Dame University, Notre Dame Indiana, 46556, USA \\
$^2$ Department of Physics, University of Oradea, Oradea, 3700, Romania 
}  
\date{\today} \maketitle 
\begin{abstract} 
We performed a self-consistent calculation of the transport properties of a $d$-wave superconductor. We used for calculations the T-matrix approximation.
The coresponding equations were evaluated numerically directly on the real
frequecy axis. We studied the $ab$-plane charge dynamics in the coherent
limit. For the $c$-axis charge dynamics, we considered both, the coherent
and the incoherent limit. We also have calculated the penetration depth in
this model.

\end{abstract}

] \makeatletter 
\global\@specialpagefalse 

\let\@evenhead\@oddhead 
\makeatother 

\section{Introduction}

Although, a large number of experimental and theoretical investigations
indicate that two dimensionality of the normal and superconducting state is
one of the key factors in high temperature superconductivity \cite{1,2}, it
remains to examine whether the cuprates are two-dimensional ($2D$) metals or
three-dimensional ($3D$) metals with strong anisotropy. The high temperature
superconductors show remarkable deviation from Fermi liquid behavior in the
normal state, such as the appearance of the pseudogap phenomena and their
related issues. The pseudogap phenomena mean the suppression of the low
frequency spectral weight without any long range order. There are enormous
studies from both experimental \cite{3} and theoretical point of view \cite
{4} in order to explain these phenomena. The direct measurements of
electronic spectrum such as $ARPES$ \cite{5} have indicated the similarity
between the pseudogap and superconducting gap while using Intrinsic
Tunneling Spectroscopy \cite{6}, it was found that the pseudogap, is
coexisting with the superconducting gap, indicating a different nature of
the two phenomena. The crucial test for superconducting origin of the gaps
is their magnetic field dependencies. Magnetic field is a strong depairing
factor and destroys superconductivity when the field exceeds the upper
critical field $H_{c2}$.

The opening of the pseudogap has drastic effect on the physical properties
of the high $T_c$ cuprates. It is found that associated with the pseudogap
the in-plane resistivity deviates from the $T$-linear behavior \cite{7} and
the $T$ coefficient of the $c$-axis resistivity changes sign, signaling
semiconductor like behavior \cite{8}. A remarkable point of the pseudogap is
that it's structure in momentum space is the same as the superconducting $%
d_{x^2-y^2}$ symmetry with continuous evolution through $T_c$\cite{9}. This
implies that the pseudogap phenomena have close connection to the
superconducting fluctuations \cite{10}, and strongly suggested that the
pseudogap is a precursor of the superconductivity \cite{11}. The main
difference between the superconducting gap and the pseudogap is that the gap
function is caused by the superconducting order, while the pseudogap is
caused by the superconducting fluctuations. The optical conductivity is one
of the quantities which most evidently display the anisotropy of the system.
Measurements of the $ab$-plane conductivity suggest that the conductance in
plane is coherent both in underdoped and overdoped regimes while the $c$%
-axis conductivity changes in character from a coherent behavior in
overdoped regime to an incoherent behavior in the underdoped regime \cite{13}%
. Theoretical investigations of the $ab$-plane and $c$-axis conductivity
both in the normal and superconducting phase can be found in \cite{14}.

In this paper we extend the self-consistent T-matrix calculation to the
superconducting state. We consider the effects of the superconducting
fluctuations on the electronic state which are the origin of the pseudogap.
In the superconducting state the T-matrix approximation includes both the
amplitude and the phase mode of the superconductor order parameter. We
investigate the transport properties which show the pseudogap phenomena. We
calculate their behavior in the superconducting state. The normal state
behavior was calculated in \cite{12}. This paper is constructed as follows.
In section II we give the Hamiltonian and explain the theoretical framework.
In section III we present the results for the spectral function $A({\bf k}%
,\omega )$ and the results for the transport properties studied in this
paper. In section IV we present the conclusions.

\section{ Theoretical Framework}

\subsection{ Model Hamiltonian}

In this section we present the theoretical framework of this paper. We
consider a microscopic model, which incorporates both strong electron
fluctuation in the $CuO$ planes and a weak interlayer coupling. The hopping
between layers is included in the following Hamiltonian: 
\begin{equation}
H=\sum_lH_l-\sum\limits_{{\bf l}}t_c\left( i\right) \left( c_{l{\bf ,}%
i,\sigma }^{+}c_{l+1,i,\sigma }+c_{l+1{\bf ,}i,\sigma }^{+}c_{l,i,\sigma
}\right)  \label{1}
\end{equation}
where $t_c$ is the hopping amplitude between layers and $c_{l,i,\sigma }(c_{l%
{\bf ,}i,\sigma }^{+})$ is the annihilation (creation) operator for an
electron within the planar site $i$, with spin $\sigma $, in layer $l$.
Within each layer we consider the following two-dimensional model
Hamiltonian which has $d_{x^2-y^2}$ symmetry superconducting ground state: 
\begin{eqnarray}
H_l &=&\sum\limits_{{\bf k,}\sigma }\varepsilon _{{\bf k}}c_{l{\bf ,k}%
,\sigma }^{+}c_{l{\bf ,k},\sigma }  \label{2} \\
&&+\sum\limits_{{\bf k,k}^{\prime }{\bf ,}\sigma }V_{{\bf k,k}^{\prime }}c_{l%
{\bf ,k},\uparrow }^{+}c_{l{\bf ,k+q},\uparrow }c_{l{\bf ,k}^{\prime
},\downarrow }^{+}c_{l{\bf ,k}^{\prime }-{\bf q},\downarrow }  \nonumber
\end{eqnarray}
where $c_{_{{\bf l,k},\sigma }}(c_{_{{\bf l,k},\sigma }}^{+})$ is the
annihilation (creation) operator for an electron with momentum ${\bf k}$,
and spin $\sigma $ in the layer $l$. The electron dispersion relation is
given by: 
\begin{equation}
\varepsilon _{{\bf k}}=-2t\left( \cos k_x+\cos k_y\right) -4t^{\prime }\cos
k_x\cos k_y-\mu  \label{3}
\end{equation}
where $t$ and $t^{\prime }$ are the nearest and the next-nearest neighbors
hopping amplitudes and $\mu $ is the chemical potential. For now on we will
consider $t$ equal to unity and $t^{\prime }=-0.5t.$ The pair interaction is
written as: 
\begin{equation}
V_{{\bf k,k}^{\prime }}=Vf_{{\bf k,k}^{\prime }}  \label{4}
\end{equation}
where 
\begin{equation}
f_{{\bf k}}=\cos k_x-\cos k_y  \label{5}
\end{equation}
is the $d_{x^2-y^2}$ wave factor. In equation (\ref{4}) $V$ is negative. The
important character of high $T_c$ cuprates is the momentum dependence of the
interlayer hopping matrix element $t_c({\bf k})$. The transfer matrix $t_c(%
{\bf k})$ obtained by the band calculation \cite{15} is expressed as: 
\begin{equation}
t_c({\bf k})=\left( cos k_x-cos k_y\right) ^2  \label{6}
\end{equation}
The large magnitude of the resistivity anisotropy in the normal state
reflects that the c-axis mean free path is shorter than the interlayer
distance, and the carriers are tightly confined to the $CuO$ planes, and
also is the evidence of the incoherent charge dynamics in the $c$-axis
direction.

\subsection{Self-Consistent T-Matrix Approximation}

In this paper we focused on the calculation of the Green function directly
on the real frequency axis \cite{16} in order to avoid the difficulties of
controlling the accuracy of the calculations used in the numerical
analytical continuations from the imaginary to the real axis by Pad\'e
algorithm \cite{17}. In the previous paper \cite{12} we presented the
self-consistent equations set, which must be solved in order to obtain the
Green function in the normal state. In this section we will present the
equations only for the superconducting state. In the superconducting state,
the self-consistent T-matrix approximation is a conserving approximation in
the sense of Baym and Kadanoff \cite{18}.

For a continuum model with local interactions, the corresponding equations
have been derived by Haussmann \cite{19} and the analogous equations for the
Hubbard model by Pedersen et. al. \cite{20}. The equations were solved in 
\cite{21} and the thermodynamics of a superconductor was also studied. A
similar model was also studied in \cite{22}.

Here we extend the self-consistent T-matrix calculation to the
superconducting state. We carry out a self-consistent calculation for the
spectral functions: 
\begin{equation}
A({\bf k},\omega )=-\frac{1}{\pi} Im G({\bf k},\omega )  \label{7}
\end{equation}
and: 
\begin{equation}
B({\bf k},\omega )=-\frac 1\pi ImF({\bf k},\omega )  \label{8}
\end{equation}
where the Green function $G({\bf k},\omega )$ is given by: 
\begin{equation}
G({\bf k},\omega )=\frac{\omega +\varepsilon _{{\bf k}}+\Sigma (-{\bf k}%
,-\omega )}{\left( \omega -\varepsilon _{{\bf k}}-\Sigma ({\bf k},\omega
)\right) \left( \omega +\varepsilon _{{\bf k}}+\Sigma (-{\bf k},-\omega
)\right) -\Delta _{{\bf k}}^2}  \label{9}
\end{equation}
and the anomalous Green function $F({\bf k},\omega )$ is given by: 
\begin{equation}
F\left( {\bf k},\omega \right) =\frac{\Delta _{{\bf k}}}{\left( \omega
-\varepsilon _{{\bf k}}-\Sigma ({\bf k},\omega )\right) \left( \omega
+\varepsilon _{{\bf k}}+\Sigma (-{\bf k},-\omega )\right) -\Delta _{{\bf k}%
}^2}  \label{10}
\end{equation}
where $\Sigma ({\bf k},\omega )$ is the retarded self-energy and $\Delta _{%
{\bf k}}$ is the order parameter in the superconducting state. We choose for
the gap function $\Delta _{{\bf k}}$ a $d_{x^2-y^2}$-symmetry form $\Delta _{%
{\bf k}}=$ $\Delta f_{{\bf k}}$. The imaginary part of the self-energy can
be expressed as: 
\begin{eqnarray}
Im\Sigma ({\bf k},\omega ) &=&f_{{\bf k}}^2\sum_{{\bf k}^{\prime }}\int
d\omega ^{\prime }\left[ f(\omega ^{\prime })+n(\omega +\omega ^{\prime
})\right] \times   \label{11} \\
&&A({\bf k}^{\prime },\omega ^{\prime })ImT_{11}({\bf k+k}^{\prime },\omega
+\omega ^{\prime })  \nonumber
\end{eqnarray}
where $f(\omega )$ and $n(\omega )$ are the Fermi-Dirac and Bose-Einstein
distributions. The real part of the retarded self-energy can be calculated
using the Kramers-Kr\"onig relation: 
\begin{equation}
Re\Sigma ({\bf k},\omega )=p.v.\int \frac{d\omega ^{\prime }}\pi \frac{%
Im\Sigma ({\bf k}^{\prime },\omega ^{\prime })}{\omega -\omega ^{\prime }}
\label{12}
\end{equation}
where $p.v.$ represent the principal value of the integral. We have to
mention that only the diagonal part of the T-matrix is taken into account
when we calculated the self-energy. Ignoring the off-diagonal part of the
T-matrix means that the anomalous self-energy is not considered. The
T-matrix is given by the following relation: 
\begin{equation}
T({\bf q},\omega )=V\left[ 1+V\Pi ({\bf q},\omega )\right] ^{-1}  \label{13}
\end{equation}
where $\Pi ({\bf q},\omega )$ is a $2\times 2$ matrix given by: 
\begin{equation}
\Pi ({\bf q},\omega )=\left( 
\begin{tabular}{ll}
$K({\bf q},\omega )$ & $L({\bf q},\omega )$ \\ 
$L^{*}({\bf q},\omega )$ & $K(-{\bf q},-\omega )$%
\end{tabular}
\right)   \label{14}
\end{equation}
The imaginary part of $K({\bf q},\omega )$ and $L({\bf q},\omega )$ can be
calculated as follows: 
\begin{eqnarray}
ImK({\bf q},\omega ) &=&\pi \sum_{{\bf k}}\int d\omega ^{\prime }f_{{\bf k}%
}^2tgh\frac{\omega ^{\prime }}{2T}\times   \label{15} \\
&&A({\bf k},\omega ^{\prime })A({\bf q-k},\omega -\omega ^{\prime }) 
\nonumber
\end{eqnarray}
\begin{eqnarray}
ImL({\bf q},\omega ) &=&\pi \sum_{{\bf k}}\int d\omega ^{\prime }f_{{\bf k}%
}^2tgh\frac{\omega ^{\prime }}{2T}\times   \label{16} \\
&&B({\bf k},\omega ^{\prime })B({\bf q-k},\omega -\omega ^{\prime }) 
\nonumber
\end{eqnarray}
The real part of the $K({\bf q},\omega )$ and $L({\bf q},\omega )$ are also
calculated using Kramers-Kr\"onig relation. The chemical potential is
determined by fixing the carriers density $n$ through the relation: 
\begin{equation}
n=2\sum_{{\bf k}}\int d\omega A({\bf k},\omega )f(\omega )  \label{17}
\end{equation}
We will consider in the following, the hole doping $\delta =1-n.$ We also
have the sum rule which must be satisfied: 
\begin{equation}
1=\sum_{{\bf k}}\int d\omega A({\bf k},\omega )  \label{18}
\end{equation}
The gap $\Delta $ can be calculated self-consistently using the gap equation
which in our notations can be written as: 
\begin{equation}
\Delta =V\sum_{{\bf k}}\int d\omega f_{{\bf k}}^2B({\bf k},\omega )f(\omega )
\label{19}
\end{equation}
Equations (\ref{17})-(\ref{19}) must be satisfied at each step in the
self-consistent calculation in order to obtain the new chemical potential
and the gap function. We have to mention at this point that we did not take
into account the anomalous self-energy at this level of calculations. The
gap function can be calculated from the equation: $1+VK({\bf 0},0)-VL({\bf 0}%
,0)=0$, which is equivalent to the gap equation and is realized in the
superconducting state. We determined the gap $\Delta _K$ using both methods
and found similar behavior as function of $T$.

For this calculation we used a Brillouin zone divided into $32\times 32$
lattice and the frequency integration was done over $1024$ points. All the
correlations and the convolutions appearing in Eqs. (\ref{11})-(\ref{19})
were done using the $FFT$ algorithm. Performing the self-consistent
calculation we obtain the spectral functions $A({\bf k},\omega )$ and $B(%
{\bf k},\omega )$. Using this spectral functions we can calculate different
physical characteristics of cuprates. We can also calculate the density of
state: 
\begin{equation}
N(\omega )=\sum_{{\bf k}}A({\bf k},\omega )  \label{20}
\end{equation}
The $ab$-plane conductivity can be calculated using the following relation 
\cite{14}: 
\begin{eqnarray}
\sigma _{ab}(\nu ) &=&-\frac{e^2}d\frac 1\omega \sum_{{\bf k}}\left[ \left( 
\frac{\partial \varepsilon _{{\bf k}}}{\partial k_x}\right) ^2+\left( \frac{%
\partial \varepsilon _{{\bf k}}}{\partial k_y}\right) ^2\right] \times 
\label{21} \\
&&\int \frac{d\omega ^{\prime }}\pi \left[ f(\omega ^{\prime }+\omega
)-f(\omega ^{\prime })\right] \times   \nonumber \\
&&\left[ A({\bf k},\omega ^{\prime })A({\bf k},\omega +\omega ^{\prime })+B(%
{\bf k},\omega ^{\prime })B({\bf k},\omega +\omega ^{\prime })\right]  
\nonumber
\end{eqnarray}
The equation (\ref{21}) for the $ab$-plane conductivity neglects the vertex
corrections and is coherent in character. For the calculation of $c$-axis
conductivity we consider both the coherent and incoherent limits. The
coherent $c$-axis conductivity is given by the relation: 
\begin{eqnarray}
\sigma _c(\omega ) &=&-de^2\frac 1\omega \sum_{{\bf k}}t_c^2({\bf k})\int 
\frac{d\omega ^{\prime }}\pi \left[ f(\omega ^{\prime }+\omega )-f(\omega
^{\prime })\right] \times   \label{22} \\
&&\left[ A({\bf k},\omega ^{\prime })A({\bf k},\omega +\omega ^{\prime })+B(%
{\bf k},\omega ^{\prime })B({\bf k},\omega +\omega ^{\prime })\right]  
\nonumber
\end{eqnarray}
In equations (\ref{21})-(\ref{22}) $d$ represents the interlayer distance.
The contribution from the incoherent process has been discussed by different
authors \cite{23}. The contribution from the tunneling process can be
written as: 
\begin{eqnarray}
\sigma _{inc}(\omega ) &=&-de^2\frac 1\omega \sum_{{\bf k,k}^{\prime }}\int 
\frac{d\omega ^{\prime }}\pi \left[ f(\omega ^{\prime }+\omega )-f(\omega
^{\prime })\right] \times   \label{23} \\
&&A({\bf k},\omega ^{\prime })A({\bf k}^{\prime },\omega +\omega ^{\prime })
\nonumber
\end{eqnarray}
We calculate also, the London penetration depth. This quantity is
proportional with the superfluid density. The in-plane penetration depth is
given by: 
\begin{equation}
\frac 1{\lambda _{ab}^2}=2\frac{e^2}d\sum_{{\bf k}}\int d\omega \left[
\left( \frac{\partial \varepsilon _{{\bf k}}}{\partial k_x}\right) ^2+\left( 
\frac{\partial \varepsilon _{{\bf k}}}{\partial k_y}\right) ^2\right] \left|
F({\bf k},\omega )\right| ^2f\left( \omega \right)   \label{24}
\end{equation}
and the $c$-axis penetration depth is: 
\begin{equation}
\frac 1{\lambda _c^2}=8de^2\frac 1\omega \sum_{{\bf k}}\int d\omega t_c^2(%
{\bf k})\left| F({\bf k},\omega )\right| ^2f\left( \omega \right) 
\label{25}
\end{equation}
In the next section we present the results obtained for the transport
properties calculated in this paper and compare the results with other
theoretical works and with experimental data.

\section{Results}

In this section we present the results obtained for the self-consistent
T-matrix approximation set of equations, introduced in the previous section.
Through out of our calculation we chose for the coupling constant $V=-5.5t.$
To study the superconducting state, we first calculated superconducting
transition temperature $T_c$. The superconducting transition temperature was
determined as the highest temperature where the set of equations (\ref{7})-(%
\ref{19}) has a non zero solution for the gap function. Near the critical
temperature the algoritm becomes unstable because of the rapid growth of
order parameter which was found also in the $FLEX$ calculations in \cite{24}%
, caused by the depairing effect due to low frequency spin fluctuations. In
our case we consider the effect of the superconducting fluctuations. The
effect is included in the retarded self-energy, which is determined
self-consistently. Using T-matrix approximation we include both the
amplitude and phase mode of the superconducting fluctuations in the
self-energy \cite{25}. The self-energy corrections are reduced in the
superconducting state because the fluctuations are reduced. The T-matrix
approximation gives a unified description for the normal state in the
pseudogap region of the underdoped cuprates and the superconducting state 
\cite{12}. The approximation is not accurate near the critical temperature $%
T_c$ because of the strong suppression of the depairing effect due to the
pseudogap, followed by rapid growth of superconducting gap below $T_c.$ The
order parameter grows more rapidly than in the $BCS$ model. We chose $\delta
=0.15$ as the hole concentration. We found the $BCS$ critical temperature $%
T_{BCS}=0.37t$ and in the case of the self-consistent calculations $%
T_C=0.22t $. In $Fig.1$ we present the order parameter $\Delta $ as function
of temperature.

\begin{figure}[tbp]
\centerline{\includegraphics[width=2.5in]{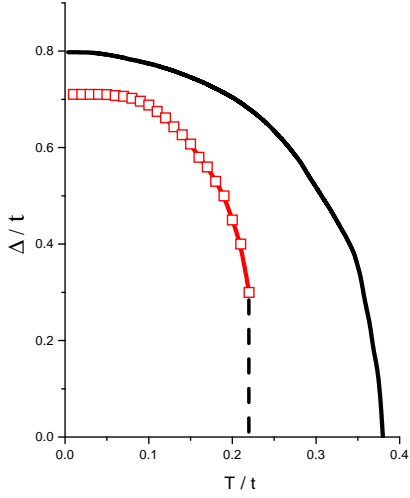}} \vspace*{3ex}
\caption{Temperature dependence of the order parameter $\Delta $. The solid
line represents the $BCS$ gap function while the symbols line represent the
self-consistent gap function. }
\label{fig:Fig1}
\end{figure}

After performing the self-consistent calculation we can calculate the
transport properties of the system. The penetration depth can be calculated
along the $ab$-plane and c-axis directions, using Eqs. (\ref{24}) and (\ref
{25}). In $Fig.2$ and $Fig.3$, we present the temperature dependence of the
penetration depth as function of temperature

\begin{figure}[tbp] 
\centerline{\includegraphics[width=2.5in]{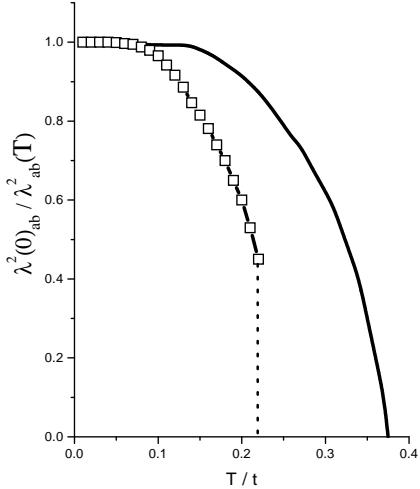}} \vspace*{3ex} 
\caption{Temperature dependence of the in-plane penetration depth in the $BCS$ model 
(solid line) and the self-consistent calculation (symbols line) } 
\label{fig:Fig2} 
\end{figure} 

We found a similar behavior as in the case of $BCS$ model. Due to the rapid
increase of the order parameter blow $T_C$ we found a rapid decrease of the
penetration depth with decreasing temperature in both $ab$-plane and $c$%
-axis directions. Recently, the penetration depth was measured as function
of temperature by different authors for $YBCO$ and $Bi-2212$ using different
techniques \cite{26}. The magnetic field penetration depth $\lambda $ of
superconductors is related to the superconducting carrier density $n_s$
divided by the effective mass $m^{*}$ as $1/\lambda ^2\propto n_s/m^{*}$. In
a recent paper Uemura \cite{37} explained the experimental data found in 
\cite{36} assuming a microscopic phase separation between superfluid and
non-superconducting fermionic carriers, similar to the superfluid $He$ films.

\begin{figure}[tbp] 
\centerline{\includegraphics[width=2.5in]{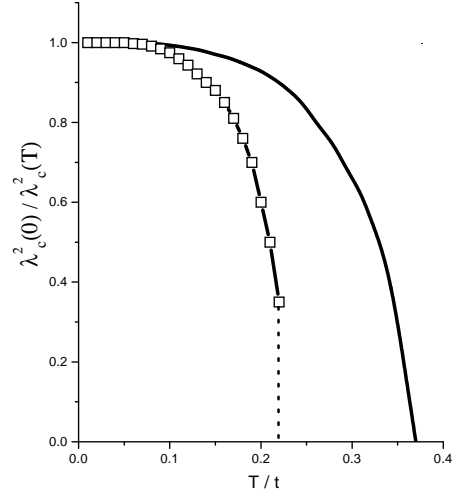}} \vspace*{3ex} 
\caption{ Temperature dependence of the $c$-axis penetration depth in the $BCS$ model 
(solid line) and the self-consistent calculation (symbols line) } 
\label{fig:Fig3} 
\end{figure} 
 
Recently, it was reported that, the photoemission spectra near the Brillouin
zone boundary, exhibit unexpected sensitivity to the superfluid density \cite
{27}, supporting the universality as function of doping and temperature of
the properties of the cuprates. We did not consider the vertex corrections
for the calculation of $F({\bf k},\omega )$. Desired improvements of the
present approach include the consideration of the vertex corrections
neglected in the present approach.

We also analyze the $ab$-plane and $c$-axis optical conductivity in the
superconducting state using the same approximation. For the calculation of $%
\sigma _{ab}$ we consider a coherent nature of the interlayer coupling. For
the $c$-axis conductivity we consider the coherent and the incoherent
coupling. Coherent coupling originates from an overlap of electronic wave
functions between planes, and in-plane momentum is conserved in interlayer
hopping. By contrast, for impurity mediated incoherent coupling, the
in-plane momentum is not conserved.

The influence of the nature of the interlayer coupling on $c$-axis
conductivity was studied by different authors \cite{14, 28}. The $ab$-plane
conductivity is mainly due to the quasiparticles near the 'cold points',
although the $c$-axis conductivity is due to the quasiparticles near the
'hot points'. Since the superconducting $d_{x^2-y^2}$ gap and the pseudogap
in the underdoped region above $T_c$ are large at 'hot points', the $c$-axis
conductivity reflects the pseudogap more clearly then the $ab$-plane
conductivity.

In $Fig.4$ we present the calculated $\sigma _{ab}$, based on Eq. (\ref{21}%
). The behavior in the superconducting state is similar with the behavior in
the normal state. 

\begin{figure}[tbp] 
\centerline{\includegraphics[width=2.5in]{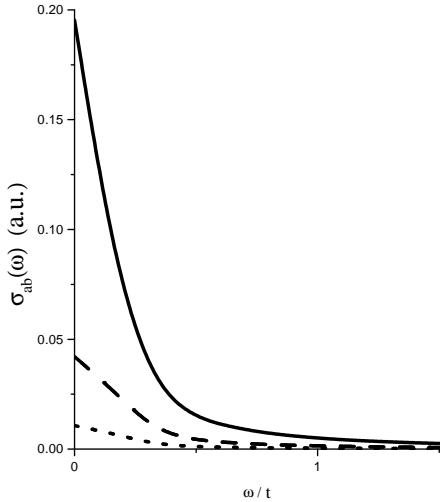}} \vspace*{3ex} 
\caption{ Frequency dependence of the $ab$-plane conductivity for different 
temperatures. The solid line corresponds to $T=0.19t$, the dashed line to $%
T=0.1t$ and the dotted line to $T=0.07t$. The critical temperature is $%
T_C=0.22t$. } 
\label{fig:Fig4} 
\end{figure} 
 
The in-plane spectrum is dominated by a Drude peak at $\omega =0$. The
intensity of the peak, increases with increasing temperature. Below $T_c$
there is no signature of superconducting gap seen in $YBa_2Cu_3O_{7-\delta }$
\cite{13}. Careful study of optical conductivity by different methods shows
that the width of the Drude peak diminished by several orders of magnitude
by decreasing the temperature, just below $T_c$ \cite{29}. In cuprates, in
contrast with $BCS$ theory, the excitations are electronic, and as the gap
develops in this excitations, the decrease in the scattering take place. The
optical properties of the cuprates, are those of a clean limit of a $BCS$
superconductor. For the calculations of the $c$-axis conductivity we
consider both the coherent and the incoherent nature of the interlayer
coupling. 

In $Fig.5$ we present the results for the $c$-axis optical conductivity in 
the case of the coherent nature of the interlayer coupling. The 
approximation assumes the independent electron propagation in each layer and 
is justified for $t_c\ll t$. In this case, the $\sigma _c$ spectrum is 
dominated by a Drude peak at $\omega =0$. This features are characteristics 
of the $d$-wave symmetry, and due to the contribution from the gap node. 

\begin{figure}[tbp] 
\centerline{\includegraphics[width=2.5in]{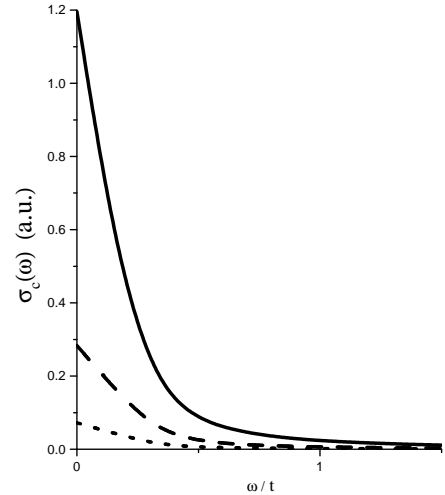}} \vspace*{3ex} 
\caption{ 
Frequency dependence of the coherent $c$-axis conductivity for different 
temperatures. The solid line corresponds to $T=0.19t$, the dashed line to $%
T=0.1t$ and the dotted line to $T=0.07t$. The critical temperature is $%
T_C=0.22t$. } 
\label{fig:Fig5} 
\end{figure} 
 
\begin{figure}[tbp] 
\centerline{\includegraphics[width=2.5in]{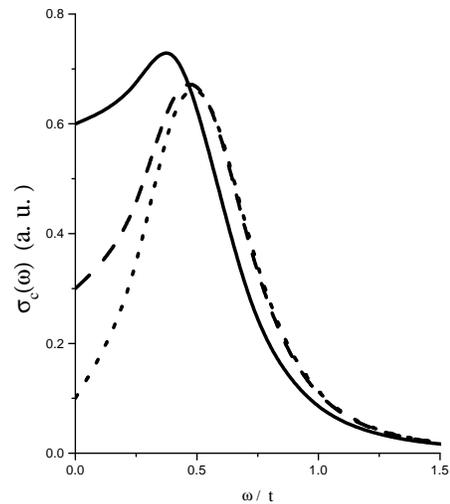}} \vspace*{3ex} 
\caption{  Frequency dependence of the incoherent $c$-axis conductivity for different 
temperatures. The solid line corresponds to $T=0.19t$, the dashed line to $%
T=0.1t$ and the dotted line to $T=0.07t$. The critical temperature is $%
T_C=0.22t$.} 
\label{fig:Fig6} 
\end{figure}

In $Fig.6$, the results for the $c$-axis conductivity are presented in the
limit of the incoherent nature of the interlayer coupling. In this case the
electronic contribution to $\sigma _c\left( \omega \right) $ does not form a
Drude peak at $\omega =0$ \cite{30}. In the superconducting state, the $c$%
-axis optical conductivity is suppressed furthermore with decreasing
temperature and shows the gap structure at low temperatures. We consider
both limit, because measurements of $c$-axis conductivity suggest that
conductance in $c$-direction is coherent in the overdoped regime \cite{31},
and incoherent in the underdoped regime \cite{13}. The coherent $\sigma _c$
was calculated using Eq. (\ref{22}) and the incoherent $\sigma _c$ using Eq.
(\ref{23}).

The results presented in $Fig.6$ are in agreement with the experimental
results obtained for the underdoped $YBa_2Cu_3O_{7-\delta }$. Here we
calculate the conductivity by neglected the vertex corrections. The vertex
correction is not important, except for the Umklapp scattering in the case
of electron correlation \cite{32}.

\section{Conclusions}

Solving the self-consistent T-matrix equations for the model presented in
the section 2 we have computed the transport properties of the system in the
superconducting state. Calculations were performed for the superconducting
order parameter of $d_{x^2-y^2}$ symmetry. The effect of the superconducting
fluctuations is included in the self-energy corrections. It is clear that
the T-Matrix approximation breaks down in the limit $T\rightarrow T_C$ since
the propagator is dramatically renormalized at the Fermi level. Our scenario
is based on resonance scattering \cite{11}. The gap function was found to
develop more rapidly than in the $BCS$ model. We calculated the London
penetration depth and the $c$-axis and $ab$-plane conductivity and compared
the results with the experimental data. The study of experimental dependence
of the penetration depth is an important problem \cite{37,36}.

Desired improvements of the present approach include the considerations of
the vertex corrections neglected in this approximation. The vertex
corrections have a significant effect in the calculation of Hall resistivity 
\cite{33} and magnetoresistance \cite{34} in high-$T_c$ superconductors.
Although, the effect was suggested to be insignificant to superconductivity 
\cite{35}, a self-consistent calculation included vertex corrections is
desired.

\end{document}